\documentclass[usenatbib]{mn2e}
\usepackage{epsfig}
\usepackage{amsmath}
\usepackage{amssymb}
\usepackage{journal_names}
\usepackage{color}
\usepackage{multirow}

\begin{document}

\title[Tidal interaction between NGC~4365 and NGC~4342]{The SLUGGS Survey: New evidence for a tidal interaction between the early type galaxies NGC~4365 and NGC~4342}
\author[Blom~et~al.~]{Christina Blom,$^{1\star}$ Duncan A.\ Forbes,$^{1}$ Caroline Foster$,^{2,3}$\\
\\
\normalfont{\LARGE Aaron J.\ Romanowsky$^{4,5}$ and  Jean P.\ Brodie$^{5}$} \\
\\
\normalsize{$^{1}$Centre for Astrophysics and Supercomputing, Swinburne University, Hawthorn, VIC 3122, Australia}\\
\normalsize{$^{2}$Australian Astronomical Observatory, PO Box 915, North Ryde, NSW 1670, Australia}\\
\normalsize{$^{3}$European Southern Observatory, Alonso de Cordova 3107, Vitacura, Santiago 19001, Chile}\\
\normalsize{$^{4}$Department of Physics and Astronomy, San Jos\'e State University,1 Washington Square, San Jose, CA 95192, USA}\\
\normalsize{$^{5}$University of California Observatories, 1156 High St., Santa Cruz, CA 95064, USA}\\
\\
$^\star$ \normalfont{Email: cblom@astro.swin.edu.au}
}

\date{\today}
\maketitle

\begin{abstract}
We present new imaging and spectral data for globular clusters (GCs) around NGC~4365 and NGC~4342. NGC~4342 is a compact, X-ray luminous S0 galaxy with an unusually massive central black hole. NGC~4365 is another atypical galaxy that dominates the $W'$ group of which NGC~4342 is a member. Using imaging from the MegaCam instrument on the Canada-France-Hawaii Telescope (CFHT) we identify a stream of GCs between the two galaxies and extending beyond NGC~4342. The stream of GCs is spatially coincident with a stream/plume of stars previously identified. We find that the photometric colours of the stream GCs match those associated with NGC~4342, and that the recession velocity of the combined GCs from the stream and NGC~4342 match the recession velocity for NGC~4342 itself. These results suggest that NGC~4342 is being stripped of GCs (and stars) as it undergoes a tidal interaction with the nearby elliptical galaxy NGC~4365. 
We compare NGC~4342 to two well-known, tidally stripped galaxies (M32 and NGC 4486B) and find various similarities. We also discuss previous claims by \citet{Bo12a} that NGC~4342 cannot be undergoing significant tidal stripping because it hosts a large dark matter halo. 
\vspace{0.2in}
\end{abstract}

\section{Introduction}

$\Lambda$CDM cosmology predicts that the giant galaxies we see in the local Universe have been built up to their present size by successive mergers of galaxies and accretion of small galaxies. Globular clusters (GCs) are very useful tracers of these merger and accretion events as they are dense and mostly robust to the violent interactions of galaxies. Additionally, several extragalactic spectroscopic studies have found that the overwhelming majority of GCs are $>10$ Gyrs old \citep{St05}. Therefore, most of the GCs that we observe locally were formed very early in the evolution of the Universe and have survived all the interactions their host galaxies were involved in. They have likely undergone several galaxy merger events, or have been stripped off smaller galaxies, and accumulated around large galaxies over time \citep{C98, To13}. They should maintain a chemical signature of the conditions they were formed under and a kinematic signature of the processes by which they were acquired by large galaxies \citep{We04}.

NGC~4365 is a giant elliptical galaxy with a redshift independent distance measurement of 23.1 Mpc \citep{FCS5} and a recession velocity of 1243 km s$^{-1}$ \citep{vSys}, placing it $\sim 6$ Mpc behind the Virgo Cluster. NGC~4365 is the central galaxy in the $W'$ group and has been noted to have rare properties for a galaxy of its luminosity. It has a kinematically distinct core and the bulk of its stars rotate along the minor axis \citep[][Arnold et al.\ 2013 in prep.]{Su95,Da01,Kr11}, while its GC system consists of three subpopulations \citep{Pu02,La05,Br05,Blom12a,Blom12b} rather than the usual two \citep{Br06,VCS9}. It has recently been studied as part of the SAGES Legacy Unifying Globulars and GalaxieS (SLUGGS\footnote[1]{http://sluggs.swin.edu.au}; Brodie et al.  in preparation) Survey.

NGC~4342 is a small S0 galaxy 20 arcmin away from NGC~4365 with a recession velocity of 751 km s$^{-1}$ \citep{Gr98} but its distance has not been measured independently of redshift. Recently, \citet{Jia12} searched the Seven Samurai Survey \citep{Fa89} and highlighted it as a local example of a high redshift superdense `red nugget' \citep{Da05,vD08} and it has long been known to have an oversized central supermassive black hole for its bulge mass \citep{Cr99}.

In the last year, \citet{Bo12a} presented a very deep $B$ filter image of NGC~4365 showing a stellar stream extending from NE of NGC~4365, across the elliptical galaxy, bridging the gap between NGC~4365 and NGC~4342, and extending further SW. This image has prompted investigation of the nature of the interaction between NGC~4365 and NGC~4342. \citet{Bo12b} found X-ray emitting hot gas around NGC~4342. This presents an interesting puzzle regarding the origin of the stellar stream. It appears as if the stellar material has been tidally stripped off NGC~4342 as it passed by NGC~4365 and what remains of NGC~4342 is now swinging back towards NGC~4365. However, the presence of X-ray emitting gas around NGC~4342 suggests a large dark matter halo that should not be present if stars have been stripped off the galaxy. 

Here we investigate the interaction between NGC~4365 and NGC~4342 using GCs, assessing what the spatial, colour and velocity distributions of the GCs around NGC~4342 and NGC~4365 indicate about the nature of the interaction between these two galaxies. In particular, is NGC~4365 stripping stars and GCs off NGC~4342 or is there another explanation?

We describe the photometric selection of GC candidates and detail our analysis of the properties of the spatial and colour distributions of the GC system in Section 2. Section 3 contains the reduction and analysis of the spectroscopic data, the kinematic selection of GCs as well as a kinematic analysis of the stream and NGC~4342 GCs. In Section 4 we investigate the effects of tidal stripping with respect to galaxy scaling relations and present a counter-argument to some of the findings of \citet{Bo12a}, before discussing the results and concluding the paper in Sections 5 and 6 respectively.

\section{Photometric Analysis}

\subsection{Imaging Data}
We use the public archival data from the MegaCam instrument on the Canada-France-Hawaii Telescope (CFHT) to identify GC candidates in the area around NGC~4342 and NGC~4365. The square degree $u$ filter image stack is centred on R.A. $=185.912^\circ$ and Decl. $=7.0562^\circ$ (J2000.0) with a total exposure time of 4240 s. The $g$ and $i$ filter image stacks are centred on R.A. $=186.133^\circ$ and Decl. $=7.2708^\circ$ (J2000.0), and have total exposure times of 3170 and 2055 s respectively. The overlap of the three image stacks covers an area of $\sim0.66$ square degrees, encompassing NGC~4365, NGC~4342 and almost all of the stellar stream \citep{Bo12a}. We use catalogues produced by the MegaPipe data reduction pipeline \citep{Gw08}, operated by the Canadian Astronomy Data Centre (CADC), to spatially match point sources found in all three image stacks. The faintest $i$ magnitude of included objects is 23.0 mag. We correct for foreground reddening, using A$_u=0.11$, A$_g=0.08$ and A$_i=0.04$, determined from Galactic 
dust maps \citep{Dustmaps}.

\subsection{Globular cluster candidate selection}

\begin{figure}\centering
 \includegraphics[width=0.47\textwidth]{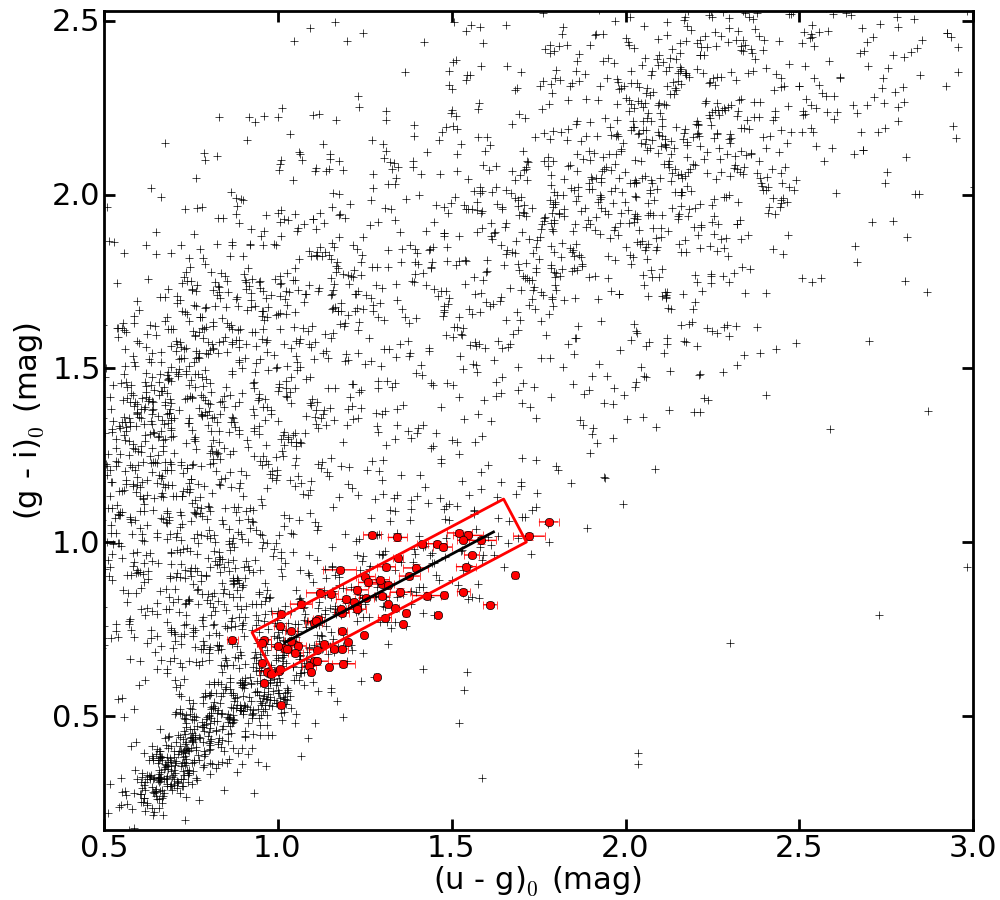}
 \caption[Colour selection of globular cluster (GC) candidates.]{Colour selection of globular cluster (GC) candidates. $(g-i)_0$ colour is plotted against $(u-g)_0$ colour. The black crosses show all point sources brighter than $i=21$ mag and red dots show the objects matched with NGC~4365 GC candidates selected from Subaru/Suprime-Cam photometry \citep{Blom12a}. The previously photometrically selected GC candidates form a linear correlation, shown with a black line. We select additional GC candidates that are consistent, within individual errors (not shown), with the red $\pm2\sigma$ box around the line.}
 \label{fig:Mcolsel}
\end{figure}

The selection of GC candidates from the matched point sources is shown in Figure \ref{fig:Mcolsel}. The figure shows $(g - i)_0$ colour plotted against $(u - g)_0$ colour for the point sources in the field-of-view, as well as the NGC~4365 GC candidates used to identify the locus of GCs in colour-colour space. These NGC~4365 GC candidates were matched in spatial coordinates between the MegaPipe catalogues and previous work using the Subaru/Suprime-Cam (S-Cam) instrument, with a $35\times27$ arcmin$^2$ field-of view \citep{Blom12a}. 
We use the S-Cam identified GC candidates to fit a line to the GC distribution, then expand the line by $\pm2\sigma$ to form a box and select candidates that overlap that box within individual errors. The spectroscopically confirmed GCs from \citet{Blom12b} are consistent with this box. Given the $i=23.0$ mag cutoff, the mean individual errors in the GC candidate sample are $\pm0.1$ mag in $(u - g)_0$ colour and $\pm0.04$ mag in $(g - i)_0$ colour. We identify a total of $1786$ GC candidates that meet these criteria.

\subsection{Spatial distribution}

\begin{figure*}\centering
 \includegraphics[width=0.98\textwidth]{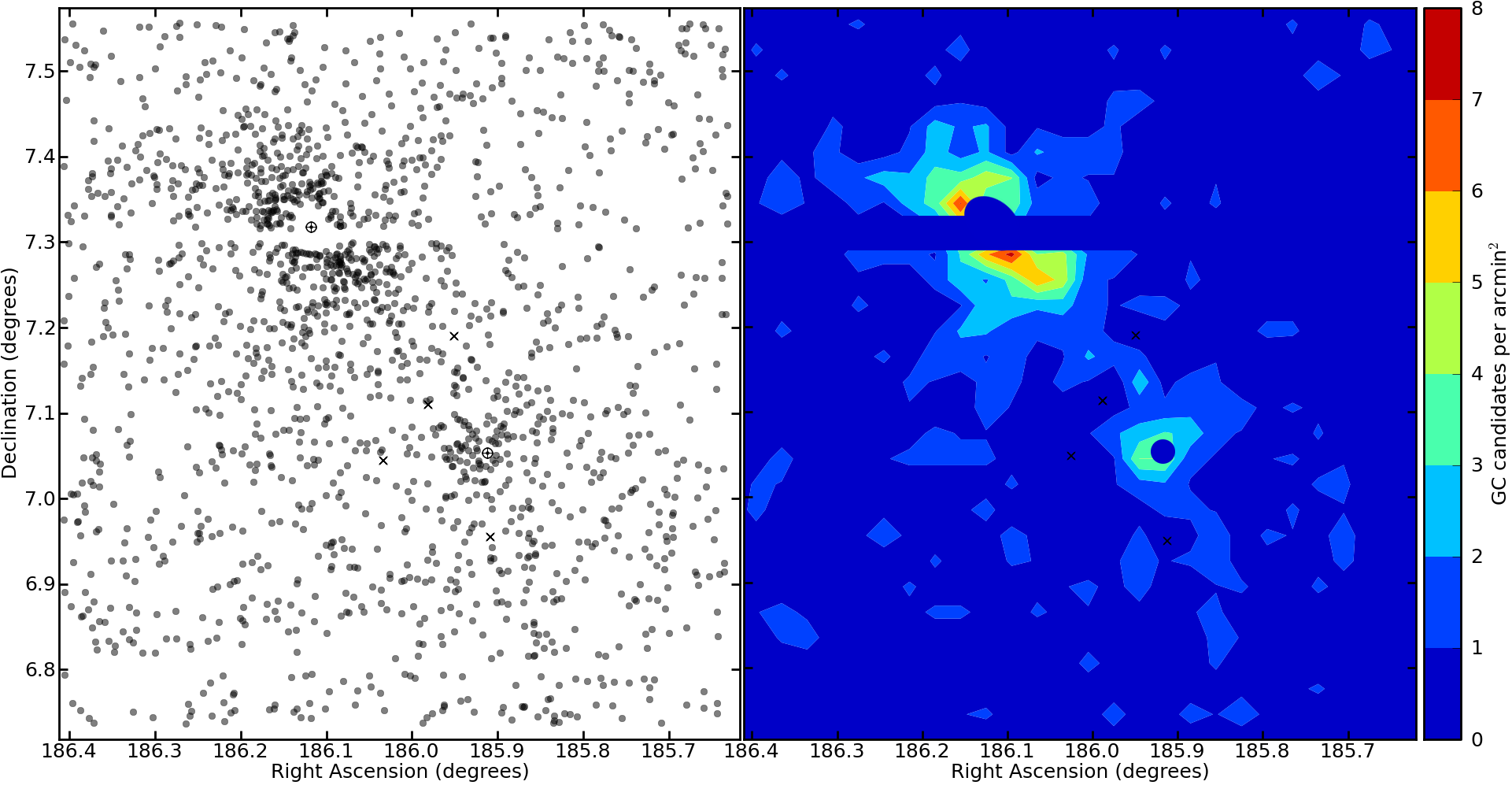}
 \caption[Spatial distribution of GC candidates around NGC~4342 and NGC~4365.]{Spatial distribution of globular cluster (GC) candidates around NGC~4342 and NGC~4365. \textbf{(Left)} Individual GC candidates are shown as grey dots and the two galaxies are shown with circles circumscribing crosses (NGC~4365 towards the top left and NGC~4342 towards the bottom right). The two-dimensional distribution of the GC candidates shows a clear overdensity around NGC~4365 and NGC~4342. There are several galaxies in the region that are as bright as NGC~4342 (marked with X's) but none show a similar GC overdensity. The horizontal void running through the centre of NGC~4365 is due to a CCD chip problem in the $u$ filter. GC candidates are not recovered in the very central regions of NGC~4365 and NGC~4342 due to the high galaxy surface brightness there. \textbf{(Right)} The two-dimensional spatial distribution of GC candidates smoothed to a resolution of 1.8 arcmin. Here we see an overdensity of GC candidates not only around NGC~4342 and NGC~4365 but also between the two galaxies and South West of NGC~4342 that is spatially coincident with the stellar stream. The spatial coincidence of the elongated GC candidate overdensity and the recently reported stellar stream \citep{Bo12a} is highly unlikely to be a chance occurrence. Both galaxy centres and the horizontal void are blocked in the smoothed image.}
 \label{fig:smoothspatial}
\end{figure*}

Given the selection of GC candidates, Figure \ref{fig:smoothspatial} shows their spatial distribution. The first panel shows the unsmoothed 2D surface density map. We see a GC overdensity corresponding to NGC~4365's GC system and we find an overdensity of GC candidates around NGC~4342. The absolute magnitude of NGC~4342 (M$_B=-18.45$ mag for a distance of $23.1$ Mpc) is comparable to that of several other galaxies superimposed on the stellar stream (i.e.\ NGC~4341: M$_B=-17.68$, NGC~4343: M$_B=-19.46$, IC 3267: M$_B=-17.7$, IC 3259: M$_B=-17.91$, assuming the same distance as NGC~4342); however we do not see a clear overdensity associated with any of them. None of the superimposed galaxies are likely to be significantly more distant than NGC~4342 (see their quoted recession velocities in Figure \ref{fig:image}). We find that NGC~4342 has an order of magnitude more GC candidates than the nearby galaxies of similar luminosity, indicating that it is significantly underluminous for its GC system.

The 2D distribution of GC candidates is smoothed to a spatial resolution of 1.8 arcmin in the second panel of Figure \ref{fig:smoothspatial}. Once the data are smoothed the GC candidate overdensity along the stellar stream is also apparent. The GC overdensity bridges the gap between NGC~4365 and NGC~4342 and extends further SW of NGC~4342, in roughly the same line as the bridge part of the GC stream. This result is robust to small variations in the magnitude cut and the spatial resolution of the smoothing. The number of GC candidates contributing to the visible stream overdensities is on the order of tens of objects and it is conceivable that such a stream structure could arise by chance. This possibility is testable if the distribution of GCs expected from both NGC~4365 and NGC~4342 was simulated and compared with the actual spatial distribution. In order to simulate the spatial distribution of the GC contribution from these two galaxies, their radial and azimuthal distributions need to be determined.

\subsubsection{Spatial properties of the GC system of NGC~4342}

\begin{figure}\centering
 \includegraphics[width=0.47\textwidth]{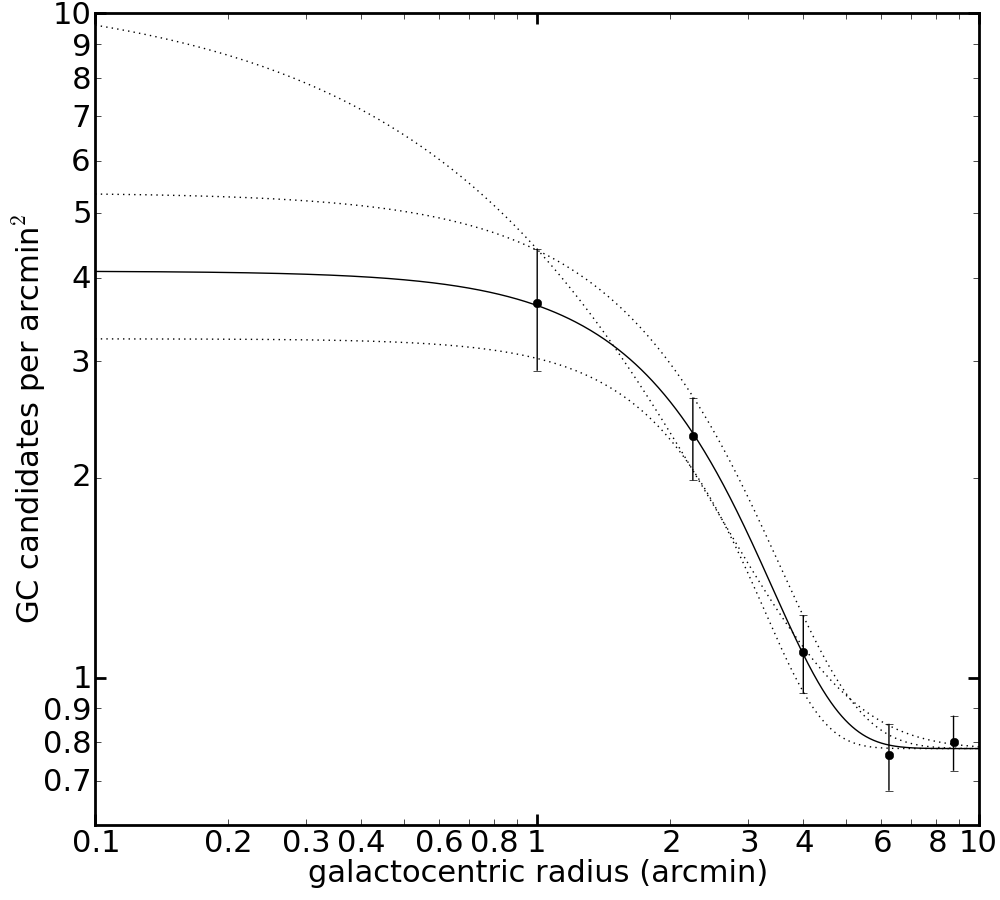}
 \caption[Radial surface density of GC candidates around NGC~4342.]{Radial surface density of GC candidates brighter than $i=23$ mag around NGC~4342 plotted against galactocentric radius. The dots with error bars show the surface density points, and the best fit S\'{e}rsic profile (with an additional background term) is plotted with a solid line. The ranges of S\'{e}rsic parameters that would still fit the data points within errors are also shown. The two dotted lines forming an envelope to the best fit line show the minimum and maximum range for the fitted parameters ($P_e$, $R_e$, $n$) respectively. The third dotted line shows a S\'{e}rsic profile where the shape parameter $n$ is 2.3 times the best fit value, consistent with all the data points. The data are well matched by the S\'{e}rsic profile fit but without data points closer than 1 arcmin from the galaxy centre it is not possible to constrain the fit well. See text for further details.}
 \label{fig:rdens}
\end{figure}

With $\sim 140$ photometrically observed GC candidates in the spatial
overdensity seen around NGC~4342 in the left panel of Figure
\ref{fig:smoothspatial} we can determine some of the quantitative
parameters of the GC distribution. The radial surface density of GC
candidates, shown in Figure \ref{fig:rdens}, is calculated in annuli
around NGC~4342. No GC candidates are detected within 0.5 arcmin from
the centre of NGC~4342 because the high starlight surface brightness
there inhibits object detection. Consequently the first radial bin
(between 0.5 and 1.5 arcmin) is possibly also affected by this
incompleteness. (We are unable to quantify this possible
  incompleteness because the original images on which object detection
  was based are not publicly available.) To ensure that there are
significant GC candidate numbers in each bin we divide the data into
only five annuli and fit a S\'{e}rsic profile added to a background
term to these. The S\'{e}rsic profile with background term ($bg$) is
given by \citep{Gr05}:
\begin{equation}
P(R)=P_e \exp\left( -b_n \left[ \left(\frac{R}{R_e}\right)^{\frac{1}{n}}-1 \right] \right)+bg
\end{equation}
where 
\begin{equation}
b_n=1.9992n-0.3271,
\end{equation}
$P_e$ is the density at the effective radius, $R_e$ is the effective radius of the GC system, and $n$ is the shape parameter of the S\'{e}rsic profile. The best fit profile is given by:
\begin{align}
P_e &= 1.68 \pm 0.19 \mathrm{\hspace{3pt}arcmin}^{-2}  \notag \\
R_e &= 2.12 \pm 0.12 \mathrm{\hspace{3pt}arcmin} \notag \\
n &= 0.50 \pm 0.12 \notag \\
bg &= 0.782 \pm 0.025 \mathrm{\hspace{3pt}arcmin}^{-2} \notag
\end{align}

The values of the background term and the effective radius of the GC
system are well constrained by the five surface density values
obtainable from the MegaCam data set. Here the value for the
background surface density will include non-GC contamination and GCs
that are spatially associated with the stellar stream rather than
NGC~4342 itself. We measure the effective radius of NGC~4342's GC
system to be $2.12\pm0.12$ arcmin ($\sim 14$ kpc) compared with the
stellar effective radius of 0.5 arcmin ($\sim 3$ kpc) \citep{RC3}. The
best fit profile is plotted on Figure \ref{fig:rdens}. The truncation
of NGC~4342 GCs at $\sim 5$ arcmin is secure and we can conclude that
the bound GC system of NGC~4342 does not extend beyond $\sim34$kpc. It
is unusual for a GC system to have a value $<1$ for the shape
parameter $n$ \citep{Rh04,Spi06,Spi08,Po13} and we also plot a
S\'{e}rsic profile where $n=1.15$, as we suspect that the
  fitted error for $n$ is underestimated due to the small dataset.
The data cannot rule out this ($n=1.15$) fit as it is consistent
within errors with every data point. To improve the constraint on the
shape parameter $n$, more GC data are needed in the central regions of
the galaxy. To probe further into the centre of NGC~4342 better
imaging and careful subtraction of the galaxy light is required. If
$n$ is indeed $<1$, the GC system of NGC~4342 is sharply truncated,
which could be an argument for tidal stripping.

\begin{figure}\centering
 \includegraphics[width=0.47\textwidth]{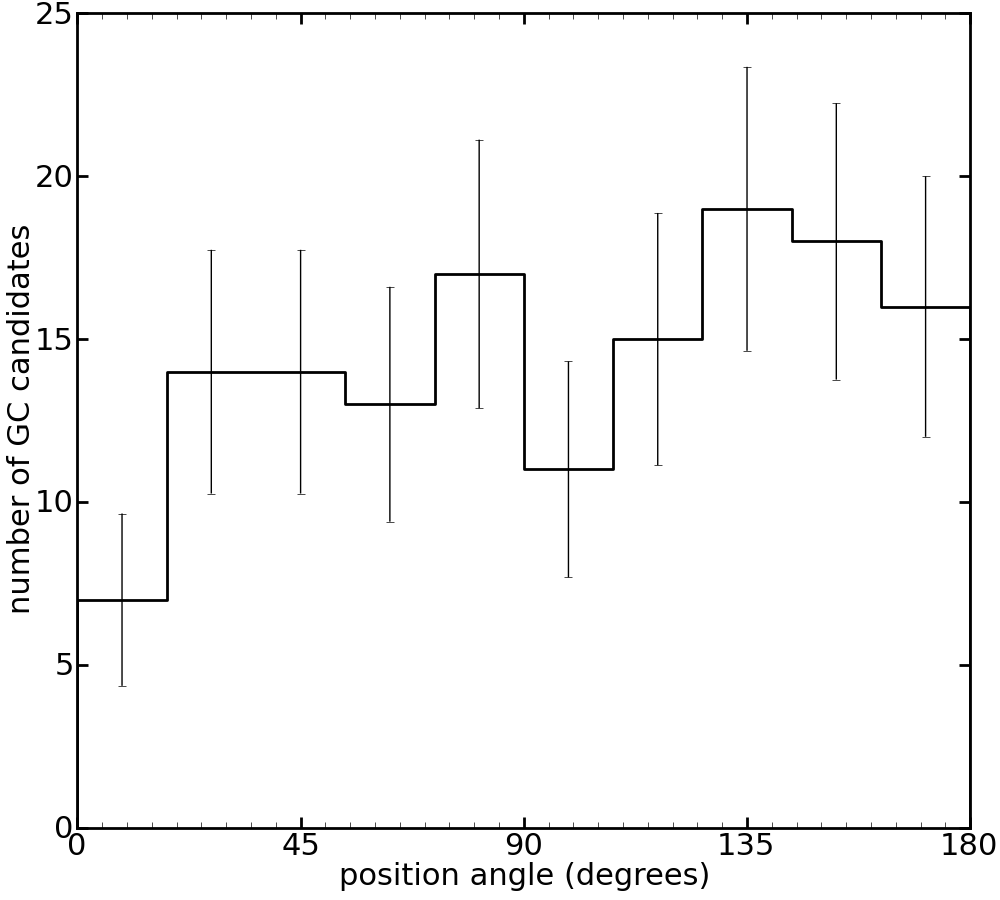}
 \caption[Azimuthal distribution of GC candidates around NGC~4342.]{Azimuthal distribution of GC candidates within 5 arcmin from the centre of NGC~4342 and brighter than $i=23$ mag. The position angle of 0$^\circ$ is directly North of NGC~4342's centre and increases anti-clockwise. The data are folded at 180$^\circ$, preserving any elliptical signature while increasing signal-to-noise. This distribution indicates that the GC system of NGC~4342 is consistent with zero ellipticity (a circular GC distribution). If the GC distribution was significantly elliptical the azimuthal distribution would show a clear sinusoidal shape with the peak of the distribution at the position angle of the elliptical distribution. The photometric position angle of NGC~4342 is $166^\circ$.}
 \label{fig:ellip}
\end{figure}

We measure the angle of each GC candidate bound to NGC~4342 (within 5
arcmin) with respect to the centre of the galaxy. Angles start
  at 0$^\circ$ directly North of the galaxy centre and progress
  anti-clockwise to 180$^\circ$ directly South of the centre. To
  increase the signal-to-noise we fold the data at 180$^\circ$ and
  plot the binned data in Figure \ref{fig:ellip}. This fold preserves
  any sinusoidal signature of ellipticity. The azimuthal distribution
of the GC candidates still bound to NGC~4342 is consistent
  with being circular. Even with azimuthal bins encompassing $18$
degrees in radius, the distribution of GC candidates does not show a
significant peak at any one position angle. It is possible that deeper
imaging, probing fainter GCs and closer to the galaxy centre would
show a small but significant ellipticity in the bound GCs. The
photometric major axis position angle of NGC~4342 is $166^\circ$ and
it would be interesting to determine if the GC distribution position
angle is consistent with that of the galaxy or the stellar stream
($\sim45^\circ$).

\subsubsection{Statistical significance of the GC overdensity on the stream}

\begin{figure}\centering
 \includegraphics[width=0.47\textwidth]{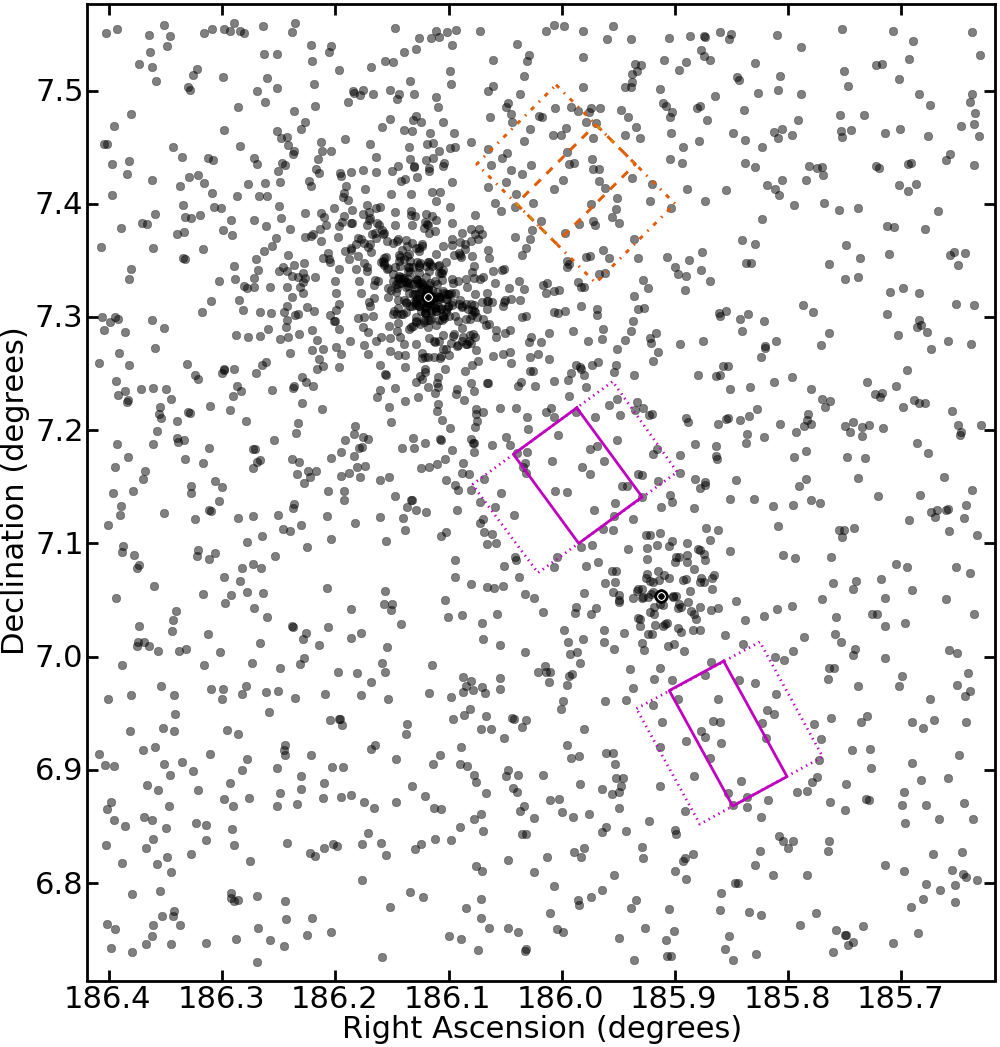}
 \caption[Simulated spatial distribution of the GCs around NGC~4365 and NGC~4342.]{An example of the 1000 simulations of the spatial distribution of the GCs around NGC~4365 and NGC~4342. Each simulated spatial distribution contains 2000 GC candidates, created by the addition of a uniform background, the S\'{e}rsic profile of NGC~4342 GCs and the spatial distribution of NGC~4365 GCs. NGC~4365's GC spatial distribution includes the S\'{e}rsic profile and azimuthal distribution fitted in \citet{Blom12a}. The solid magenta regions mark the areas of stream overdensity between the two galaxies and SW of NGC~4342. These stream overdensities can be seen in Figure \ref{fig:smoothspatial}.  The dotted magenta off-stream regions on either side of the solid boxes are used to compare off-stream densities. The dashed orange region marks the position of an overdensity to the NW of NGC 4365 also visible in Figure \ref{fig:smoothspatial} and the dash-dotted region marks the comparison area.}
 \label{fig:mock}
\end{figure}

We generated 1000 simulated GC spatial distributions around NGC~4365 and NGC~4342 of which Figure \ref{fig:mock} is an example. These distributions are the addition of a uniform background sampling added to a sampling from the spatial distributions of both NGC~4365 and NGC~4342. Each simulated spatial distribution contains 2000 objects. This number was conservatively chosen to be higher than the number of objects in the observed distribution ($1786$) because there are voids in the observed spatial distribution. The background was set at a density of 0.5 arcmin$^{-2}$ across the entire area. This value was chosen to visually match the object density far from NGC~4365, NGC~4342 and the stream on the CFHT/MegaCam observations. The background value of 0.78 measured in Section 2.3.1 is likely to include stream GCs as well as non-GC contaminating objects. The rejection method was used to extract random positions according to the spatial distribution functions. We used the S\'{e}rsic profile and azimuthal distribution fitted by \citet{Blom12a} to simulate the radial and azimuthal distribution of NGC~4365 and the previously fitted S\'{e}rsic profile for NGC~4342. The S\'{e}rsic profile was noted to be highly uncertain in the inner regions of NGC~4342, but the fitted profile is sufficient for these purposes as we are primarily concerned with the outer regions of both galaxies' spatial distributions. No significant differences are seen in the outer regions of simulated spatial distributions when the value of the S\'{e}rsic profile $n$ parameter is greater than 1. Figure \ref{fig:mock} also shows the specific areas which were compared to the observed spatial distribution.

\begin{figure}\centering
 \includegraphics[width=0.47\textwidth]{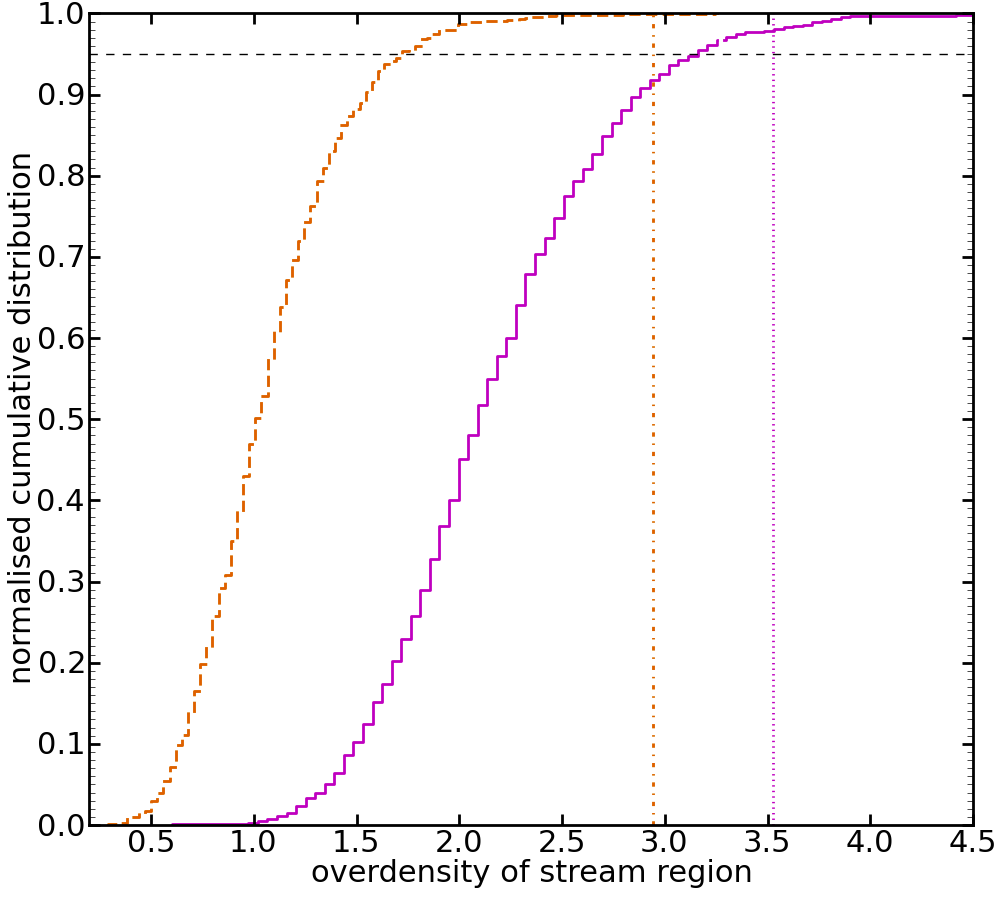}
 \caption[Calculated stream overdensities for 1000 simulated spatial distributions.]{Cumulative distribution of the calculated stream overdensity for 1000 simulated GC spatial distributions. The solid magenta line shows the on-stream density divided by the off-stream density for the stream region between the two galaxies and SW of NGC~4342 added together. The dashed orange line shows the same overdensity parameter for the region NW of NGC 4365. The dash-dotted orange and dotted magenta lines show the overdensity measured in the same regions for the observed GC candidate spatial distribution. The horizontal black dashed line shows the $95^{th}$ percentile. The intersection of the vertical measured overdensity lines and cumulative distribution lines fall well above the $95^{th}$ percentile. The overdensity NW of NGC~4365 is possibly associated with the background galaxy NGC~4334.}
 \label{fig:cumul_mock}
\end{figure}

The overdensity in the two stream areas was calculated by dividing the density in the stream defined areas by the density in the off-stream areas alongside (see Figure \ref{fig:mock}) and then added together. For the observed spatial distribution we calculated a summed stream overdensity of 3.53. We also calculated the overdensity of the region NW of NGC~4365 and found the overdensity there to be 2.94. The observed overdensities are compared with the distribution of overdensities in Figure \ref{fig:cumul_mock}. A summed overdensity equal to or greater than the observed value on the stream between NGC~4365 and NGC~4342 as well as on the stream beyond NGC~4342 occurs by chance only 2.2 percent of the time (2.3$\sigma$ in a normal distribution).  An overdensity in the regions NW of NGC~4365 occurs by chance only 0.1 percent (3.3$\sigma$) of the time. The overdensity NW of NGC~4365 is spatially coincident with the background galaxy NGC~4334 (V = 4354 km s$^{-1}$) and it is likely that the GC candidate overdensity there is associated with that galaxy, although spectroscopic observations would be required to test that further.

The GC candidate distribution along the stream is highly unlikely to occur by chance and is best explained by tidal stripping of GCs from one galaxy by another. 


\subsection{Colour distribution}
\begin{figure}\centering
 \includegraphics[width=0.47\textwidth]{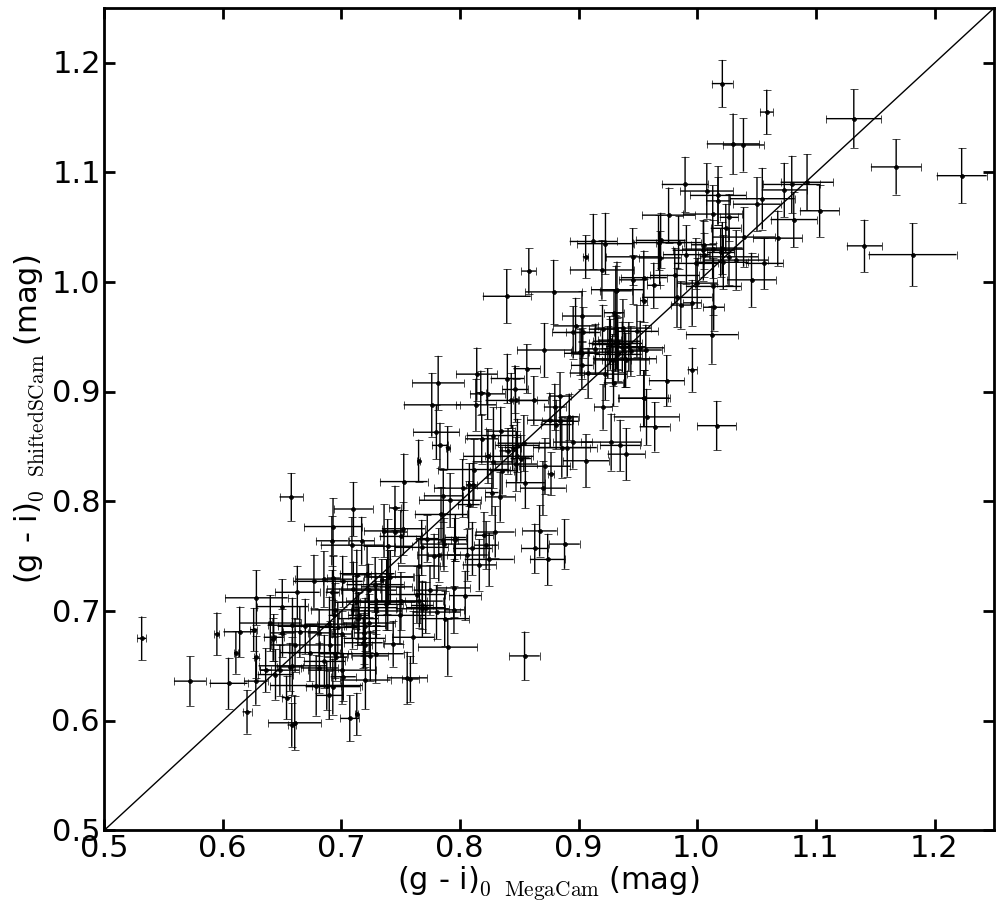}
 \caption[Scaling between Subaru/Suprime-Cam and CFHT/MegaCam colours.]{Scaling between Subaru/Suprime-Cam $(g'-i')_0$ and CFHT/MegaCam $(g-i)_0$ colours. Points with error bars are the objects with photometry from both S-Cam and MegaCam and MegaCam $i$ magnitude brighter than $23$ mag. The black line shows a one-to-one correlation between the $(g-i)_0$ colours. The S-Cam colours have been shifted by $-0.102$ mag and the comparison between shifted S-Cam and MegaCam colours is consistent with a one-to-one correlation over the GC colour range.}
 \label{fig:colshift}
\end{figure}

We have calculated the transformation from $(g'-i')_0$ in the SDSS photometric system (used for the S-Cam photometry) to $(g-i)_0$ in the MegaCam photometric system. The data used for the calculation and the resulting transformation are shown in Figure \ref{fig:colshift}. We find a shift of $-0.102$ mag from the SDSS to MegaCam colours that is not strongly dependent on $(g-i)_0$ colour over the colour range of GC candidates.

\begin{figure}\centering
 \includegraphics[width=0.47\textwidth]{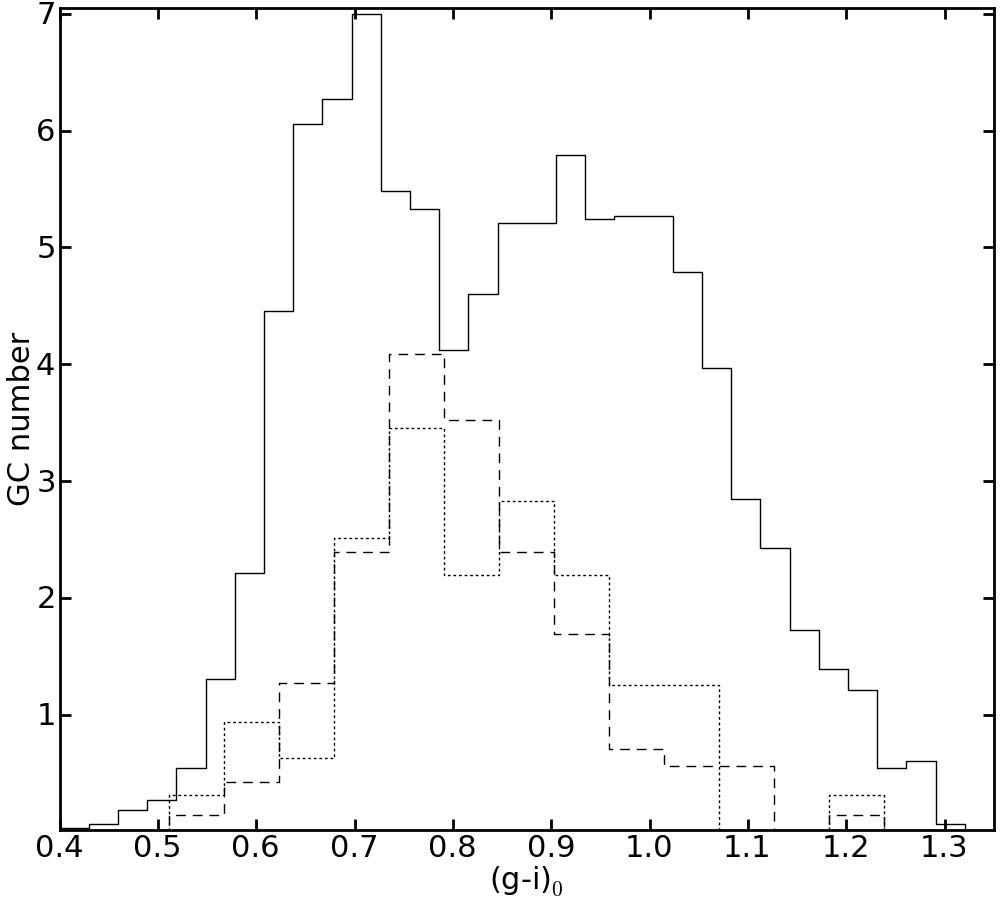}
 \caption[Colour distribution of NGC~4342 and stream GC candidates.]{Histogram of $(g-i)_0$ GC colours for NGC~4365, NGC~4342 and the stellar stream. The colour distribution of GC candidates within 5 arcmin of NGC~4342 is plotted with a dashed line and the colour distribution of candidates that overlap with the stream regions as defined in Figures \ref{fig:smoothspatial} and \ref{fig:mock} is plotted with a dotted line. The colour distribution of NGC~4365 is scaled to 33 percent of its height and plotted with a solid line. It has been shifted from Subaru/Suprime-Cam $(g'-i')_0$ \citep{Blom12a} to CFHT/MegaCam $(g-i)_0$ colours and scaled down for comparison. The colour distribution of NGC~4342's GC system is indistinguishable from the colour distribution of the GC stream and both lack the red GCs seen in the colour distribution of NGC~4365's GC system.}
 \label{fig:coldist}
\end{figure}

Comparing the colour distributions of GC candidates associated with
NGC~4365, NGC~4342 and the stellar stream (see Figure
\ref{fig:coldist}) we see a significant population of red (metal-rich)
GC candidates present in NGC~4365's GC system that is absent from the
colour distribution of either the stream or NGC~4342. We do not probe
the central regions of NGC~4342 where red GCs are likely to be found
and if the GCs are tidally stripped from the outskirts of NGC~4342 we
also do not expect metal rich GCs in the stream. There is no
significant difference between the GC colour distribution of the
stream population and that of NGC~4342, whereas the blue colour of
both stream and NGC~4342 GC candidates are not a good match for the
colour of the blue GC candidates of NGC~4365. This supports the
  theory that the stream GCs are actually GCs from the outer parts of
  NGC~4342's GC system that have been stripped off the galaxy during a
  tidal interaction with NGC~4365.

\section{Spectroscopic Analysis}

\subsection{Spectroscopic Data}
Three slitmasks were observed on 2012, April 17 with the DEep Imaging Multi-Object Spectrograph (DEIMOS) on the Keck II telescope \citep{Fa03}. The slitmasks were positioned along the stream South West of NGC~4365 and around NGC~4342. GC candidates within each slitmask were prioritized for observation based on their brightness. 
The slitmasks were each observed for a total of 2 hours, split over 4 observations of 30 min each. The median seeing was 1.2 arcseconds. For all 3 masks, the slits were 1 arcsecond wide and a $1200$ l/mm grating, centred on $7800$ \AA, was used in conjunction with the OG550 filter. This setup \citep{Po13} enables observations from $\sim 6550 - 8900$ \AA$\,$ with a wavelength resolution of $\sim 1.5$ \AA. The Calcium II triplet (CaT) absorption features (8498, 8542 and 8662 \AA) are within the observed wavelength range up to recession velocities of $\sim8$\,$200$ km s$^{-1}$. This includes NGC~4342 \citep*[V = 751 km s$^{-1}$, ][]{Gr98} and NGC~4365 \citep[V= 1243 km s$^{-1}$][]{vSys}. We use these absorption features to determine the line-of-sight velocity for each candidate GC.

\begin{figure}\centering
 \includegraphics[width=0.47\textwidth]{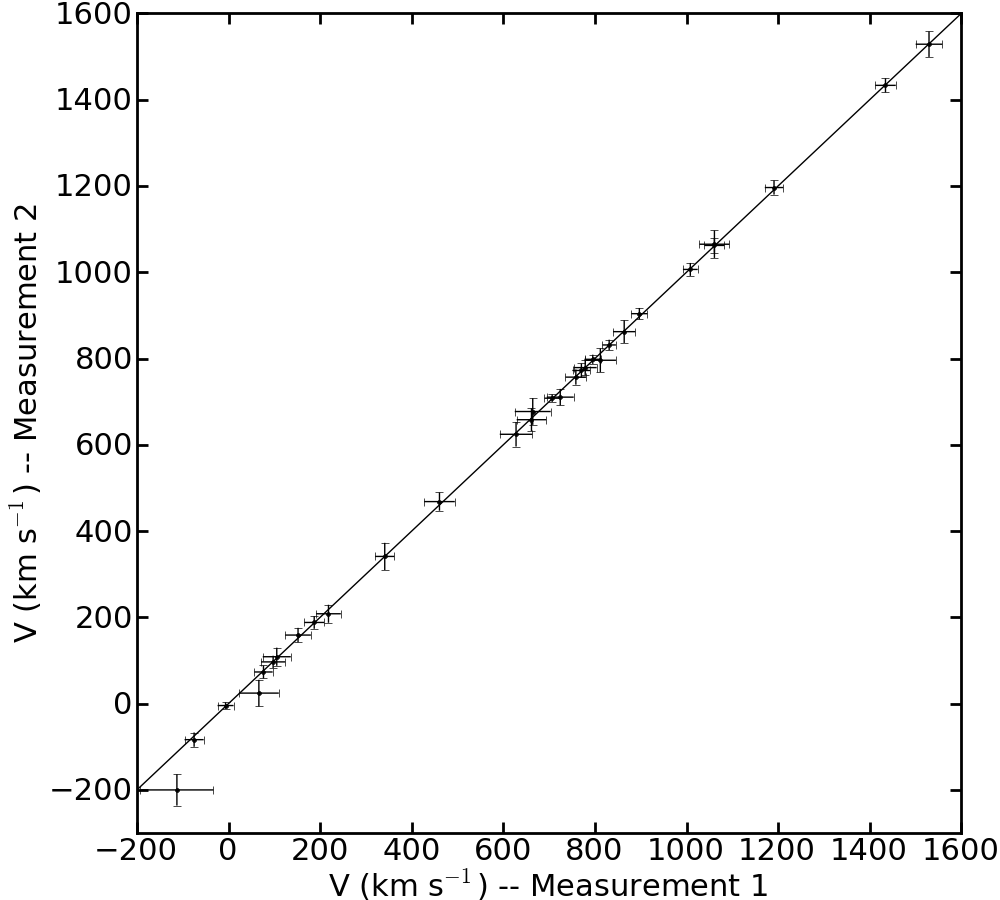}
 \caption[Comparison of independent velocity measurements.]{Comparison of independent velocity measurements from DEIMOS spectra. Velocities are plotted with error bars on a one-to-one relationship line. All the velocities are consistent with each other except the lowest velocity. We exclude this object from further analysis and measure an rms dispersion about the one-to-one line of $8.8$ km s$^{-1}$.}
 \label{fig:meascomp}
\end{figure}

The data were reduced using a modified version of the \textsc{deep}2 (Deep Extragalactic Evolutionary Probe 2) galaxy survey data reduction pipeline \citep[\textsc{idl spec2d}][]{spec2d1,spec2d2}. The pipeline uses dome flats, NeArKrXe arc lamp spectra and sky light visible in each slit to perform flat fielding, wavelength calibration and local sky subtraction, respectively. The object spectrum was cross correlated (using the \textsc{iraf} task \textsc{rv.fxcor}) with the CaT features in 13 stellar template spectra observed with the same setup. The velocity of the object is calculated by taking the mean of the 13 cross correlation values and the uncertainty on the velocity is calculated by adding in quadrature the standard deviation of the  velocities and the mean of the errors on those velocities (as determined by \textsc{rv.fxcor}). The process of velocity determination was independently repeated by a second investigator and results are shown in Figure \ref{fig:meascomp}. We only include a velocity measurement 
in further analysis, if it has a reliable velocity (as determined by both investigators) and the measured velocities are consistent within the measured uncertainties (i.e.\ the 1$\sigma$ error bar of one measurement overlaps the other velocity measurement). The root-mean-squared dispersion between the two independent measurements is $8.8$ km s$^{-1}$.

\subsection{Velocity selection}

\begin{figure*}\centering
 \includegraphics[width=0.98\textwidth]{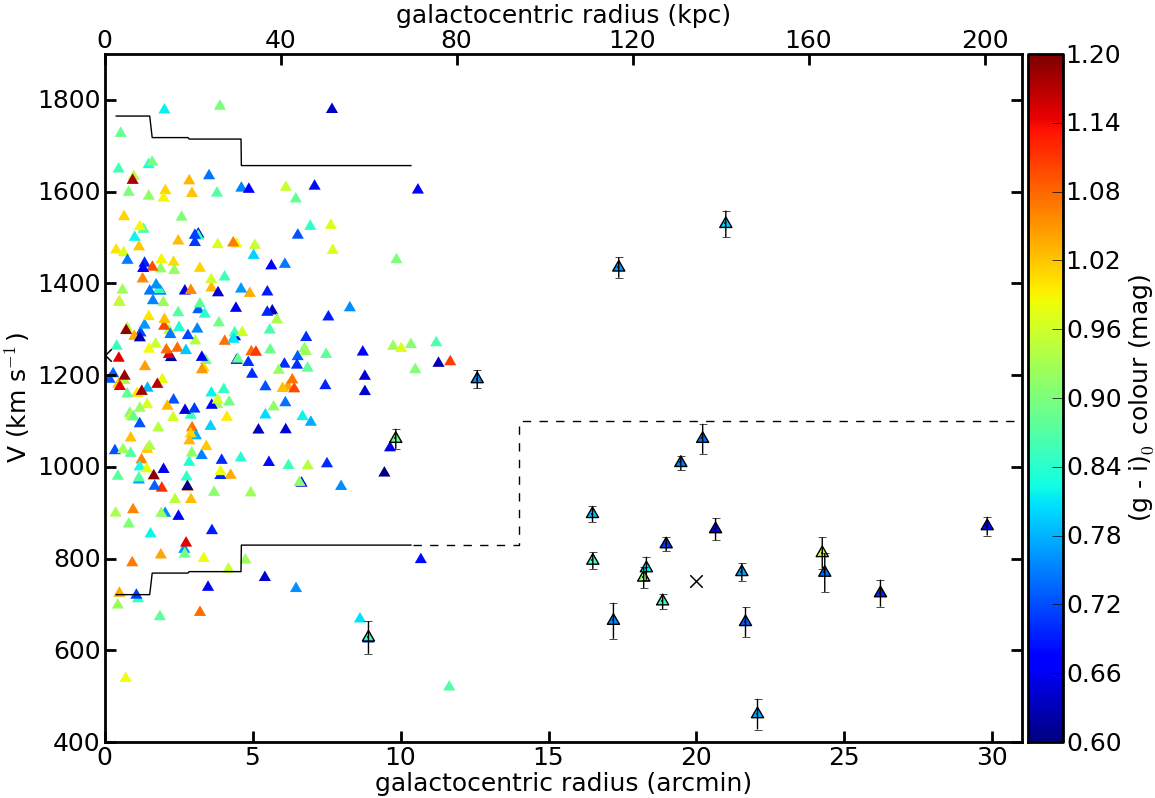}
 \caption[Phase-space diagram of GC velocity as a function of distance from NGC~4365.]{Phase-space diagram of globular cluster line-of-sight velocity as a function of distance from NGC~4365. New GC velocities presented here are shown as triangles with black outlines and error bars while previously published GC velocities \citep{Blom12b} are shown without. The colour of each symbol indicates its $(g-i)_0$ colour. NGC~4365 and NGC~4342 are marked with crosses at 0 arcmin, V = 1243 km s$^{-1}$ and $\sim$20 arcmin,  V = 751 km s$^{-1}$ respectively. The solid line shows the $2\sigma$ envelopes of the NGC~4365 GC velocity distribution and the dashed line indicates the adopted boundary, below which we classify objects as low velocity stream GCs. The 31 stream and NGC~4342 GCs have a similar mean velocity to NGC~4342. The $(g-i)_0$ colours of low velocity GCs are mostly blue with some green GCs but no red GCs. Red GCs are only found in the central regions of NGC~4365. The stream GCs are likely tidally stripped from the original GC system of NGC~4342.}
 \label{fig:velrad}
\end{figure*}

If the GC candidates on the stream and around NGC~4342 originate from the same system, we expect their recession velocities to be similar. A superposition of NGC~4342 GCs on a population of stream GCs that have been stripped from another, now completely disrupted galaxy, might show a large offset in recession velocity.
 
We find 33 objects with measurable recession velocities between -200 and 1600 km s$^{-1}$ but assume the 11 objects with velocities smaller than 400 km s$^{-1}$ to be stars in our own Galaxy. The remaining 21 GCs are plotted in Figure \ref{fig:velrad} showing GC recession velocity against galactocentric radius from NGC~4365. We no longer refer to these as GC candidates because their recession velocity has been confirmed. The new velocities presented in this work scatter about the recession velocity of NGC~4342 (V = 751 km s$^{-1}$). The velocities of GCs around NGC~4365 are also shown along with the $2\sigma$ dispersion envelopes as a function of galactocentric radius. As highlighted in \cite{Blom12b}, there is a grouping of low velocity GCs that are also consistent with NGC~4342's recession velocity.

\begin{table}\centering
\begin{tabular}{|c|c|c|c|c|c|}
\hline
\textbf{I.D.}	&	\textbf{R.A.}	&	\textbf{Decl.}	&	$\mathbf{V}$ 	&	$\mathbf{V_{err}}$	& $\mathbf{(g-i)_0}$\\
	&  \multicolumn{2}{|c|}{(degrees)} & \multicolumn{2}{|c|}{(km s$^{-1}$)} & (mag)\\ \hline \hline
1	&	185.9401	&	7.031459	&	1060.96	&	32.67 & 0.739	\\
2	&	185.9213	&	7.035125	&	863.96	&	24.33 & 0.666	\\
3	&	185.9598	&	7.045965	&	706.69	&	16.77 & 0.846	\\
4	&	185.9558	&	7.045967	&	831.14	&	14.73 & 0.708	\\
5	&	185.9771	&	7.048600	&	758.41	&	22.65 & 0.911	\\
6	&	185.9753	&	7.082502	&	795.31	&	17.96 & 0.861	\\
7	&	185.9512	&	7.099172	&	897.06	&	17.68 & 0.786	\\
8	&	186.0573	&	7.182025	&	627.83	&	35.51 & 0.854	\\
9	&	185.7894	&	6.944254	&	870.57	&	20.42 & 0. 674	\\
10	&	185.8825	&	6.988903	&	811.86	&	34.09 & 0.953	\\
11	&	185.8480	&	7.014668	&	769.10	&	41.91 & 0.741	\\
12	&	185.9185	&	7.019106	&	770.51	&	19.25 & 0.773	\\
13	&	185.8866	&	7.040393	&	661.51	&	32.29 & 0.727	\\
14	&	185.9483	&	7.063918	&	778.85	&	25.04 & 0.814	\\
15	&	185.7881	&	7.030769	&	724.31	&	29.35 & 0.696	\\
16	&	185.8933	&	7.083244	&	1008.32	&	16.11 & 0.748	\\
17	&	185.8355	&	7.081957	&	460.64	&	33.42 & 0.775	\\
18	&	185.8873	&	7.147331	&	664.27	&	39.24 & 0.758	\\
19	&	186.0902	&	7.243450	&	796.64	&	11.37 & 0.925	\\
20	&	186.1140	&	7.328690	&	538.85	&	7.39 &  0.984 \\
21	&	186.0164	&	7.152190	&	519.93	&	9.97	&	0.870 \\
22	&	186.0573	&	7.182080	&	624.60	&	6.26 &	0.659 \\
23	&	186.0640	&	7.224360	&	734.53	&	5.15 &	0.760 \\
24	&	186.1108	&	7.315610	&	698.58	&	15.57 & 0.923	\\
25	&	186.1011	&	7.325980	&	712.22	&	12.39 & 0.845	\\
26	&	186.1203	&	7.300140	&	720.09	&	8.61 &	0.712 \\
27	&	186.1341	&	7.368620	&	682.49	&	14.93 & 1.077	\\
28	&	186.0563	&	7.251730	&	758.64	&	12.42 & 0.642	\\
29	&	186.1403	&	7.175740	&	668.30	&	4.22 &	0.808 \\
30	&	186.1637	&	7.145680	&	797.62	&	5.46 &	0.680 \\
31	&	186.0988	&	7.293170	&	673.24	&	4.55 &	0.877\\ \hline
\end{tabular}
\caption[Recession velocities for GCs around NGC~4342 and in the stream.]{Positions, recession velocities and photometric colours (CFHT/MegaCam) for the 31 GCs associated with NGC~4342 and the stream. GCs numbered 1 to 18 are presented for the first time in this work and GCs numbered 19 to 31 are low velocity outliers previously presented in \citet{Blom12b}. }
\label{tab:veldata}
\end{table}

Four of the 22 GCs with measured velocities are more likely to be associated with NGC~4365 than NGC~4342 and the stream. Two have significantly higher velocities than NGC~4342 ($>1400$ km s$^{-1}$) and another two have reasonably high velocities ($>1050$ km s$^{-1}$); all are radially consistent with NGC~4365's GC system. We associate 18 new GC recession velocities with the sample of 13 stream GCs identified in \cite{Blom12b} giving a total of 31 GCs associated with NGC~4342 and the stream. The positions, recession velocities and photometric colours of these GCs are listed in Table \ref{tab:veldata}. We conclude that GCs associated with NGC~4342 and the stream are indistinguishable in velocity.

\subsection{GC system kinematics}

\begin{figure*}\centering
 \includegraphics[width=0.98\textwidth]{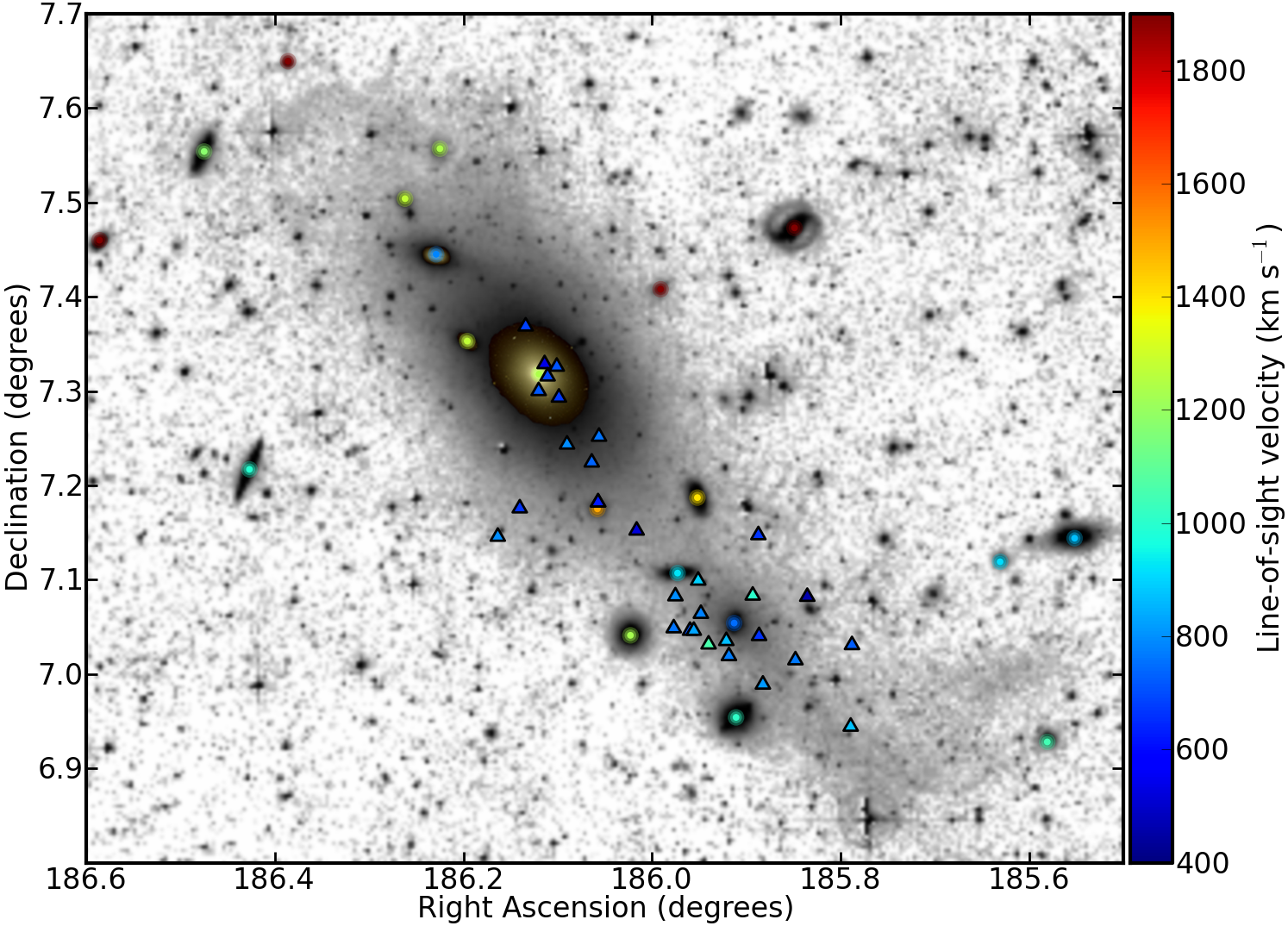}
\caption[Deep optical image of NGC~4365 and other nearby galaxies in the $W'$ group.]{Deep optical imaging of the giant elliptical NGC~4365 and other nearby galaxies in the $W'$ group, which reveals a $\sim$ 300 kpc long stream of stars \citep{Bo12a}. Globular clusters (GCs) in the stream are represented by triangles and galaxies by circles, with the colour of the points denoting their recession velocity (velocities $>$ 1900 km s$^{-1}$ are assigned  the reddest colour). The GCs have a mean velocity consistent with the recession velocity of NGC~4342 (located at R.A. $\sim 185.9^\circ$ and Dec. $\sim 7.05^\circ$) indicating that they were tidally stripped along with some stars from NGC~4342. The other galaxies in the $W'$ group  have recession velocities intermediate between those of NGC~4342 and NGC~4365 (i.e. NGC~4341: V = 922 km s$^{-1}$, NGC~4343: V = 1014 km s$^{-1}$, IC 3267: V = 1231 km s$^{-1}$, IC 3259: V = 1406 km s$^{-1}$).}
\label{fig:image}
\end{figure*}

The GCs that lie on the stream have a mean recession velocity of $682\pm22$ km s$^{-1}$ and a velocity dispersion of $83\pm16$ km s$^{-1}$. In Figure \ref{fig:image} these 14 stream GCs as well as 17 GCs more closely associated with NGC 4342 are plotted on an image of the stellar stream. The recession velocities of the GCs and nearby galaxies are marked by the colour of the points (i.e. NGC~4341: V = 922 km s$^{-1}$, NGC~4343: V = 1014 km s$^{-1}$, IC 3267: V = 1231 km s$^{-1}$, IC 3259: V = 1406 km s$^{-1}$). We note that NGC~4342 is the only galaxy in the area that has a low enough recession velocity to be consistent with the recession velocities of the GCs around it and extending along the stream. We see some indication of a possible velocity gradient along the stream but it is not possible to disentangle velocity shear along the GC stream from the effect of one-sided selection in velocity space, because NGC~4365's GC system overlaps the GCs along the stream. Overall, the GC kinematics are consistent with a cold stream associated with NGC 4342.

The GCs close, and most likely bound, to NGC 4342 have a mean recession velocity of $790\pm40$ km s$^{-1}$ and velocity dispersion of $139\pm37$ km s$^{-1}$. If three GCs with $2\sigma$ velocities are excluded from the sample, the calculated velocity dispersion decreases to $\sim 75\pm23$ km s$^{-1}$.

\section{Effects of Tidal stripping}

In this section we compare NGC~4342 to the locations of NGC~4486B and M32 on supermassive black hole (SMBH) and galaxy scaling relations. NGC~4486B and M32 are both well-known to be tidally stripped \citep{Fa73,Gr02,Choi02}. In the case of M32 a giant stream of stars was detected by \citet{Ib01} which connects it (in projection) with M31 and another satellite galaxy, NGC~205. Deep imaging of the Virgo cluster core by \citet{Mi05} reveals that NGC~4486B lies in a region of substantial intracluster light and stellar streams, although no individual stream of stars can be easily assigned to NGC~4486B. 


All three galaxies are known to deviate from the SMBH vs.\ spheroid mass relation and are relatively compact galaxies with high central velocity dispersions. Their key properties are summarised in Table \ref{tab:prop}. Here we calculate spheroid (bulge) masses from their total K band magnitudes assuming that M/L$_K$ = 0.81 \citep{Bo12a}. SMBH masses are scaled by distance, from the original source, assuming a linear dependence. NGC 4486B is located in the Virgo cluster at a distance of 16.5 Mpc \citep{FCS5}. Its projected separation from central giant elliptical NGC 4486 (M87) is $\sim$35 kpc. According to \citet{Gu09} the SMBH mass determined by \citet{Ko97} is somewhat uncertain. M32 is a close satellite of M31, for which we use a distance of 820 kpc. We note that M32 is the prototype of the compact elliptical (cE) class, which until recently included only half a dozen objects \citep{Chil07}.


\begin{table} \centering
\begin{tabular}{lcc}
\hline
{\bf Property} & {\bf Value} & {\bf Reference}\\
\hline
{\bf NGC 4342} & S0 & 1\\
Distance & 23.1 Mpc & 2\\
$\sigma_0$ & 241$\pm$10 km/s & 3\\
R$_e$ & 672 pc & 4a\\
R$_e$ (bulge) & 96 pc & 4b\\
B$_T^0$ & 13.37 mag & 1\\
M$_B$ & --18.45 mag & --\\
K & 8.9 mag & 4a\\
K (bulge) & 10.3 & 4b\\
Stellar mass & 2.5 $\times$ 10$^{10}$ M$_{\odot}$ & 2\\
Stellar mass (bulge) & 6.9 $\times$ 10$^{9}$ M$_{\odot}$ & 2\\
$[Fe/H]$ & 0.25 dex & 6\\
Black hole mass & 4.6$^{+2.5}_{-1.5}$ $\times$ 10$^8$ M$_{\odot}$ & 7\\
Globular clusters & 1200 $\pm$ 500 & 8\\
\hline
{\bf NGC 4486B} & cE0 & 1\\ 
Distance & 16.5 Mpc & 9\\
$\sigma_0$ & 291$\pm$25 km/s & 10\\
R$_e$ & 180 pc & 11\\
B$_T^0$ & 14.26 mag & 1\\
M$_B$ & --16.83 mag & --\\
K & 10.09 mag & 5\\
Stellar mass & 4.3 $\times$ 10$^{9}$ M$_{\odot}$ & 2\\
$[Fe/H]$ & 0.13 dex & 12\\
Black hole mass & 6.2$^{+3}_{-2}$ $\times$ 10$^8$ M$_{\odot}$ & 10\\
\hline
{\bf M32} & cE2 & 1\\
Distance & 820 kpc & 11\\
$\sigma_0$ & 76$\pm$10 km/s & 11\\
R$_e$ (bulge) & 100 pc & 11\\
B$_T^0$ (bulge) & 9.23 mag & 11\\
M$_B$ (bulge) & --15.34 mag & --\\
Stellar mass (bulge) & 8.0 $\times$ 10$^{8}$ M$_{\odot}$ & 13\\
$[Fe/H]$ & 0.0 dex & 14\\
Black hole mass & 2.9$^{+0.6}_{-0.6}$ $\times$ 10$^6$ M$_{\odot}$ & 15\\
\hline
\end{tabular}
\caption{Galaxy Properties. References are: (1) = \citet{RC3}; (2) = this work, see text; (3) = \citet{Fa89}; (4a) = Vika (priv. comm.); (4b) = Vika et al. (2012); (5) = \citet{2MASS}; (6) = \citet{vdB98}; (7) = \citet{Cr99}; (8) = \citet{Bo12b}; (9) = \citet{FCS5}; (10) = \citet{Ko97}; (11) = \citet{Chil07}; (12) = \citet{San06}; (13) = \citet{HR04}; (14) = \citet{Ro05}; (15) = \citet{Ve02}.}
\label{tab:prop}
\end{table}

\begin{figure*}\centering
\includegraphics[width=0.98\textwidth]{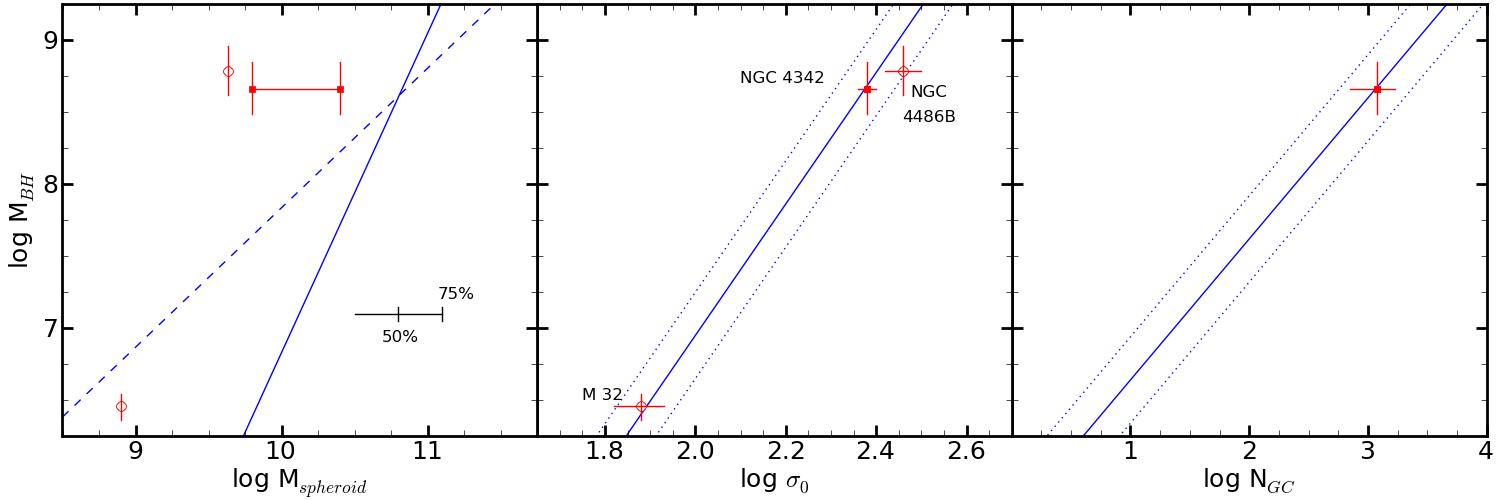}

\caption{Supermassive black hole (SMBH) scaling relations. SMBH
  scaling relations (solid lines) from the literature are shown: {\bf
    (Left)} spheroid mass in solar masses for core-Sersic (dashed) 
and Sersic (solid) profiles from Scott et al. (2013); 
{\bf (Centre)} central velocity dispersion in
  km s$^{-1}$ from \citet{Gr12} for non-barred S\'{e}rsic galaxies;
  {\bf (Right)} total number of globular clusters (GC) from
  \citet{HH11}. A representative intrinsic scatter of $\pm$ 0.3 dex
  for the velocity dispersion and GC scaling relations are 
shown by the dotted lines. NGC 4342 is shown by filled squares, with two 
spheroid masses representing the range in bulge mass. M32 and NGC 4486B are 
shown by open circles. 
The sample galaxies are labelled in the middle panel (with B for NGC 4486B). 
In the
  left panel the horizontal bar with vertical ticks shows the effect
  of 50 percent and 75 percent stellar mass loss from the original
  galaxy.  All three galaxies are consistent with the central velocity
  dispersion scaling relation (and total number of globular clusters
  in the case of NGC 4342), whereas the galaxies deviate from the
  spheroid mass relation in the direction of substantial stellar mass
  loss due to tidal stripping.}
\label{fig:bh}
\end{figure*}

The left panel of Figure \ref{fig:bh} shows the SMBH mass vs.\ spheroid
mass. 
For a galaxy undergoing tidal stripping of its outer stars, the luminosity
will decrease, while the central properties will remain largely
unchanged \citep{Be92,Choi02}. In particular, we expect its SMBH mass
and inner density profile to be unchanged, while its total stellar mass is reduced.
Although NGC 4342 is currently consistent with a single S\'{e}rsic profile
(which is unlikely to have been modified by tidal
stripping) we also include the core-S\'{e}rsic scaling relation 
from \citet*{Sc13}. 
We show two extreme data points for NGC 4342, i.e. 
the small (R$_e$ = 96 pc) bulge and the maximal bulge (the whole galaxy).  
NGC 4324, along with M32 and NGC 4486B, lies to the left of the 
standard Sersic relation to a lower spheroid mass for a given SMBH
mass. (We note that M32 has a bulge that is consistent with a Sersic profile;  
Graham 2002.) 
In other words these galaxies all have a much higher SMBH mass
as a fraction of their spheroid mass than the typical fraction of a
percent for early-type galaxies \citep[e.g.][]{Gu09,Gr12}. 
The effects of 50 percent and 75 percent 
stellar mass loss due to tidal stripping are shown.

Figure \ref{fig:bh} also shows two other tight SMBH scaling relations
-- the stellar velocity dispersion near the centre of the galaxy
\citep{Gr12} and the total number of globular clusters associated with
a galaxy \citep{HH11}. For galaxies undergoing tidal stripping, the
  central velocity dispersion ($\sigma_0$) should be largely unchanged
  \citep{Be92}. Globular clusters (GCs) are fairly robust to galaxy
  interactions and mergers \citep[their old ages indicating that many
    have survived a Hubble time, e.g.][]{St05}. So although they will
  be stripped from the host galaxy \citep[e.g.][]{Bek03}, a large area
  census should recover the original system population. Thus the {\it
    total} number of original GCs (N$_{GC}$) will also be unaffected
  by tidal stripping if a full census can be made. We expect the
  galaxies to be consistent with these scaling relations even if they
  have undergone tidal stripping.  All three galaxies are fully
  consistent with the SMBH vs.\ velocity dispersion relation. In
    the case of NGC 4342, which has an estimate of its total original
    GC system \citep{Bo12b}, it is also perfectly consistent with the
    GC scaling relation. Although the GC systems of the other
  galaxies have been studied, there is no estimate available for the
  {\it total} number of GCs in the original GC system.

\begin{figure}\centering
\includegraphics[width=0.49\textwidth]{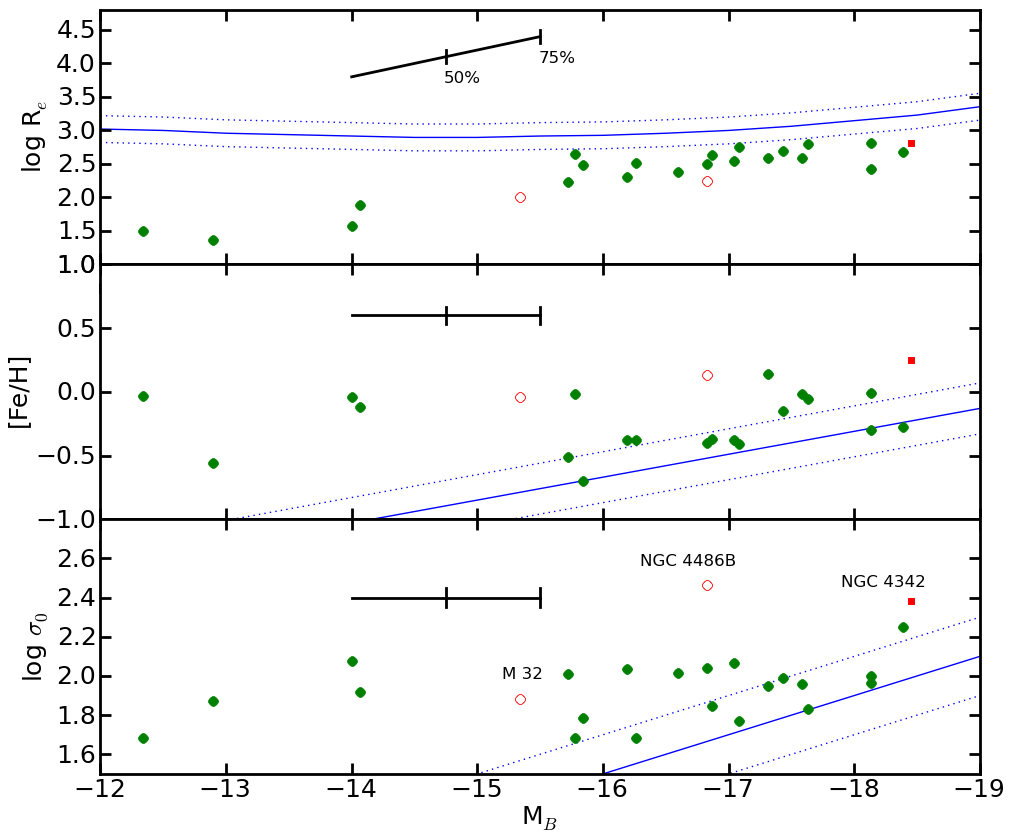}
\caption{Galaxy scaling relations. Early-type galaxy scaling relations with B band luminosity (solid lines) from the literature are shown: {\bf (Top)} effective radius in parsecs from \citet{Gr11Rev}; {\bf (Middle)} central metallicity from \citet{Chil08}; {\bf (Bottom)} central velocity dispersion in km s$^{-1}$ from \citet{MG05}. A representative scatter of $\pm$ 0.2 dex for each scaling relation is shown by the dotted lines. Sample galaxies are shown by filled red circles and labelled in the lower panel. The location of compact elliptical galaxies (thought to be tidally-stripped) from the literature are shown by green filled circles. The bars with vertical ticks shows the effect of 50 percent and 75 percent stellar mass loss and size reduction due to tidal stripping. Tidal stripping will reduce the luminosity and the effective radius  but the central velocity dispersion and central metallicity will remain largely unchanged. The galaxies are consistent with substantial stellar mass loss from the distribution of normal early-
type galaxies.}
\label{fig:mb}
\end{figure}

Further support for the tidal stripping hypothesis can be gained by comparing the galaxy stellar properties with those of normal galaxies and galaxies classified as compact ellipticals \citep[which are generally thought to be the result of tidal stripping;][]{Chil07}. Figure \ref{fig:mb} shows three key galaxy scaling relations as a function of B band galaxy luminosity, i.e. central velocity dispersion, central metallicity and effective radius. This figure shows that the three galaxies deviate from the standard scaling relations. 

The figure also shows vectors representing the expected effects of 50 percent and 75 percent tidal stripping, assuming that size is reduced by the same proportions as per \citet{Be92} and that the central values of velocity dispersion and metallicity are unaffected. For the three subplots, each galaxy is consistent with having had their luminosity, stellar mass and size reduced by $\sim$50-75 percent due to the effects of tidal stripping. Applying the stripping vectors to the current locations of the galaxies in Figure \ref{fig:mb} suggests that the original progenitor galaxies were drawn from a large range in luminosity. It would clearly be of interest to obtain SMBH masses for a large sample of compact galaxies \citep[see][]{vdB12} to determine whether they lie systematically offset from the SMBH vs.\ spheroid mass relation in the sense as the three galaxies presented here.

The tidal stream around NGC~4365 detected by \citet{Bo12a} shows an extensive (300 kpc) stellar stream with NGC 4342 lying near the centre of its SW arm. The stream has an integrated luminosity that is similar to NGC 4342 \citep{Bo12a}. Since some stream light will be below the detection threshold, the stream mass is probably more than 50 percent of the current inferred mass of NGC 4342. If the stream has been tidally-stripped from NGC~4342 then the stellar mass loss is comparable to the amounts suggested by Figures \ref{fig:bh} and \ref{fig:mb}. 

\citet{Bo12a} argued against tidal stripping of NGC~4342 for two
reasons. The first is that a spheroid mass based on the K band surface
brightness fits of Vika et al. (2012) indicated that over 90 percent
of the galaxy stars would have to have been removed to explain its
location in the SMBH vs.\ spheroid mass relation. \citet{Vi12} fitted
bulge and disk components. Their resulting bulge effective radius
R$_e$ was 0.86 arcsec, with a bulge mass of 6.7 $\times$ 10$^{9}$
M$_{\odot}$. However, the bulge size quoted by \citet{Vi12} was only
barely resolved given their seeing and sampling, and seems more akin
to a nucleus given its physical size of 96 pc. A single component
Sersic fit by Vika (priv. comm.) also gives a good fit to the K band
data with R$_e$ = 6 arcsec (similar to that measured by
\citealt{Fa89}, which we adopt here).  The {\it total} K
band magnitude gives a stellar mass of 2.5 $\times$ 10$^{10}$
M$_{\odot}$ (similar to the value used by \citealt{McC13} in their
recent large sample SMBH analysis). Thus we have two extreme values for the spheroid mass of NGC 4342, 
i.e. using the same small bulge as \citet{Bo12a} or the total galaxy as used by \citealt{McC13}
(with new photometry from \citealt{La13a} lying in between).
These extremes are shown in Figure \ref{fig:bh} and both values deviate strongly from  
standard SMBH vs.\ spheroid mass relations.  
 NGC 4342 is consistent with 50-75 percent mass loss if the spheroid mass lies between the two extreme values given. 
  A similar mass loss is suggested by the scaling relations shown in \ref{fig:mb}, and this is comparable to the estimated 
  stellar mass in the stream. Although NGC 4342 may not have been stripped of 90 per cent of its mass, such high rates of mass loss may be 
  appropriate for NGC~4486B and M32.



The second reason that \citet{Bo12a} argued against tidal stripping is
that dark matter in galaxy haloes has a lower binding energy on average than the
stars and would generally be stripped first. Without a dark matter
halo the galaxy would have difficulty retaining the hot X-ray emitting
gas as observed (a point that will be examined further in the next Section).
On the other hand, simulations suggest that stars can be stripped
without the removal of {\it all} dark matter. From hydrodynamical simulations
of small disk galaxies with $\Lambda$CDM-like haloes being stripped within a
Virgo cluster-like potential, \citet{Go08} concluded that {\it ``Even
  when the entire [baryonic] disk is tidally stripped away, the
  nucleus stays intact and can remain dark matter dominated even after
  severe stripping.''} 
  In another case, \citet{Li11} simulated the mass stripped from
small galaxies (subhaloes) located within the larger dark matter halo
of a massive galaxy. They found that after a billion years, tidal stripping
had indeed removed most of the dark matter, but 30 percent remained
along with 60 percent of the original stellar mass (i.e. 40 percent of
the stellar mass was stripped away). This work suggests that while
  most of the original dark matter is lost when half of the
  stars have been stripped away, a significant
  fraction (10 percent) of the dark matter can remain and would be likely to reside
  in the very central parts of the galaxy. 

  Despite these intriguing counterexamples, it may strain credibility to suppose
  that the stars in NGC~4342 have been almost completely stripped beyond a radius
  of $\sim$~2--3~kpc, while retaining a substantial dark matter halo out to at least $\sim$~5--10~kpc
  with no sign of truncation.
 However, there is a precedent in the mechanism of resonant 
 stripping, which is predicted to remove stars from the centres of dwarf galaxies 
 while leaving their dark haloes intact \citep{DO09}. In the \citet{DO09} simulation of a small galaxy interacting with its hundred times larger neighbour, $\sim$80 per cent of the stars were removed with little loss of dark matter. The simulations indicate that most of the stars would be removed from the galaxy's disk, therefore it is likely, in this scenario, that NGC 4342 had a very extended disk that is now almost completely disrupted by tidal stripping. This also appears to be the case for M32 which reveals an `outer disk' at large radii \citep{Gr02, Choi02}.
 For NGC~4342 orbiting NGC~4365, the condition for resonance would imply a pericentric distance of $\sim$~10--20~kpc. 
To investigate such a possibility further, and to more generally resolve the question of
whether or not hot gas and dark matter could be retained after tidal stripping, a
tailored simulation is needed. This hydrodynamical simulation would need to include, at least, hot gas, stars and GCs embedded within 
the dark matter subhalo around NGC 4342 which is interacting with NGC 4365 within the overall dark matter and hot
gas halo of the W$^{'}$ group. 

Observational searches for dark matter around other tidally stripped galaxies could also be useful.
In the case of M32,  \citet{How13} recently 
concluded that a dark matter halo was required to bind a small number of observed
high velocity stars  (with the caveat that they may not be 
in equilibrium). Thus despite the evidence for tidal stripping 
of stars from M32 \citep{Gr02,Choi02} it may still contain some 
dark matter.

\section{Discussion}


The properties of the GC system around NGC~4342 suggest that the galaxy is being tidally stripped. We find a distribution of GCs peaked at the position of NGC~4342. Detailed analysis of the GC spatial distribution reveals a GC stream overlapping with NGC~4365's GC system (NE of NGC~4342) and extending a roughly equal distance SW of NGC~4342. The $(g-i)_0$ colour distribution of the GCs close to NGC~4342 is indistinguishable from that of GCs in the stream and significantly different from that of NGC~4365's GC colour distribution. This suggests that the stream GCs have been drawn from the outskirts of NGC~4342's GC system. In addition to the similarity of their spatial and colour properties, the recession velocities of stream and NGC~4342 GCs ($\mathrm{V}=742\pm23$ km s$^{-1}$) match the recession velocity measured from NGC~4342 starlight (V = 751 km s$^{-1}$) very well.

\citet{Bo12a} reported the presence of a stellar stream extending from the NE corner of NGC~4365 across the galaxy, through NGC~4342 and further SW. This stellar stream aligns well with the GC stream. 
Given the spatial and kinematic evidence, as well as the fact that NGC~4342 is a much less luminous galaxy (M$_B=-18.45$ mag) than NGC~4365 (M$_B=-21.3$ mag) it appears that NGC~4365 is tidally stripping NGC~4342 as it moves through the $W'$ group. 

\citet{Bo12b} found X-ray evidence of hot gas (and ram pressure stripping of the gas)
surrounding NGC~4342. They inferred a massive dark matter halo under the assumption of 
hydrostatic equilibrium for the hot gas not affected by ram pressure stripping. 
Thus despite obtaining a mass profile out to $\sim$~40~kpc, they relied on the information within only $\sim$~10~kpc,
and considered the equilibrium assumption still questionable in the $\sim$~5--10~kpc region. 
Within the $\sim$~2--5~kpc region where the mass profile should be more reliable, the dark matter component is
inferred to be dominant, although we note that in other galaxies, X-ray based mass estimates have been found to be too high
by factors of $\sim$~2--3, even when the X-ray gas appears relatively relaxed
\citep{Ro09,St11,Na13}.

If the inferred X-ray mass is accurate, and NGC~4342 has a massive dark matter halo that extends far beyond
the central luminous component, then it becomes difficult to explain a tidal stripping scenario,
as discussed in the previous Section.

If the stars have instead been stripped off NGC~4365 and are in the process of accreting onto NGC~4342 we expect GCs from the outskirts of NGC~4365 to be accreting onto NGC~4342 as well. Given the measured recession velocities of NGC~4365, NGC~4342 and the stream GCs, this interpretation of the stream GCs originally coming from NGC~4365 is highly unlikely. 



Further support for the tidal stripping scenario comes from investigation of NGC~4342 in the context of galaxy scaling relations and comparison to well known tidally stripped galaxies. NGC~4342 shows an offset from the black hole mass v.s. spheroid mass as well as effective radius/metallicity/central velocity dispersion vs.\ B band luminosity scaling relations in the same direction but to a lesser degree than NGC~4486B and M32. The scaling relation analysis shows it to be consistent with 50-75 percent stellar mass loss.

Finally, it is possible that the GCs and the stream came from a fully disrupted (now unseen) galaxy. If interaction with NGC~4342 has completely disrupted this galaxy and both its stars and GCs are now being accreted onto NGC~4342, there is no need for stars to be stripped from NGC~4342 before its dark matter halo. It is not far-fetched to assume a galaxy with an almost identical GC colour distribution to NGC~4342, since many galaxies have GC systems with similar colour distributions \citep{VCS9}, but it might at first seem to be somewhat more problematic to also assume that such a galaxy had exactly the same recession velocity as NGC~4342. However, many of the galaxies around NGC~4342 have similar recession velocities and recent observations have shown that small galaxies accreting along streams or filaments might not be as uncommon as previously thought \citep{Ib13}. Nonetheless, the fully-disrupted galaxy scenario does not explain why the stellar properties of NGC 4342 match those of the stripped galaxies, M32 and NGC~4486B. The simplest interpretation is that the GCs and stars
in the stream originated in NGC 4342. 


\section{Conclusions}

We find an overdensity of globular clusters (GCs) centred on the S0 galaxy NGC~4342 that is spatially coincident with the stellar stream crossing NGC~4365, as reported by \citet{Bo12a}. The photometric colours of the stream GCs match the colours of the GCs around NGC~4342 and their measured recession velocities ($\mathrm{V}=742\pm23$ km s$^{-1}$) match the measured recession velocity of NGC~4342 itself (V = 751 km s$^{-1}$). The evidence from these observations suggests that stars and GCs from NGC~4342 are being tidally stripped by the larger galaxy NGC~4365, situated at the centre of the $W'$ group. When a wider array of stellar properties of NGC~4342 are considered, including black hole mass, metallicity and effective radius, it appears very similar to the well-known tidally stripped galaxies, NGC~4486B and M32.
This finding is in tension with the conclusion that NGC~4342's stars and GCs have {\it not} have been stripped because it still hosts a large dark matter halo as inferred from observations of hot X-ray emitting gas \citep{Bo12b}. Tailored simulations of NGC 4342, that include gas, stars and GCs within a dark matter halo are needed to reconcile all the evidence. 

\section{Acknowledgements}
We thank the reviewer for helpful comments on the manuscript and thank P.\ Nulsen and B.\ Matthews for stimulating discussions. We also thank M.\ Smith, V.\ Pota, C.\ Usher, S.\ Kartha and N.\ Pastorello for support during the preparation of this manuscript. The data presented herein were obtained at the W.M. Keck Observatory, which is operated as a scientific partnership among the California Institute of Technology, the University of California and the National Aeronautics and Space Administration. The Observatory was made possible by the generous financial support of the W.M. Keck Foundation. The analysis pipeline used to reduce the DEIMOS data was developed at UC Berkeley with support from NSF grant AST-0071048. This research used the facilities of the Canadian Astronomy Data Centre operated by the National Research Council of Canada with the support of the Canadian Space Agency. CF acknowledges co-funding under the Marie Curie Actions of the European Commission (FP7-COFUND). This work was supported in part by NSF grants AST-0909237 and AST-1211995 as well as ARC 
Discovery grant DP130100388. DAF was supported by ARC grant DP130100388.

%
%
\bibliographystyle{mn2e}
\bibliography{ref}

\end{document}